# Cu:ZnO deposited on porous ceramic substrates by a simple thermal method for photocatalytic application

D. Bouras[a], A. Mecif[a], R. Barillé[b], A. Harabi[c], M. Rasheed[b], A. Mahdjoub[d], M. Zaabat[a]

[a] Laboratory of Active Components and Materials, Larbi Ben M'Hidi University, Oum El Bouaghi 04000, Algeria.

[b] MOLTECH-Anjou, Université d'Angers/UMR CNRS 6200, 2 Bd Lavoisier, 49045 Angers, France.

[c] Ceramics Laboratory, Mentouri University, Constantine 25000, Algeria.

[d] Laboratoire des Matériaux et Structures des Systèmes Electromécaniques et leur Fiabilité (LMSSEF) Université d'Oum El Bouaghi BP 358, Algérie.

**Abstract**

Thin films of undoped zinc oxide (ZnO) and doped ZnO with copper were deposited on ceramic pellets made from abundant local clay materials with an addition of zirconia ($ZrO_2$). The thin layers were prepared by a thermal method using an autoclave. The resulting structural and morphological properties of the products were studied in order to determine the effectiveness of their photocatalytic activities. X-ray diffraction, Scanning Electron Microscopy, energy-dispersive X-ray spectroscopy and UV-visible spectrophotometry were used in this goal. An application test was carried out to quantify the photocatalytic degradation of a toxic organic dye – Orange II (OII) by these samples under UV light. In similar conditions, the layers deposited on the ceramics with addition of zirconia have shown better performances than those without zirconia. This result can be related to the porosity created when the zirconium oxide reacted with $SiO_2$ of the clay. The open porosity observed on this type of substrate allows a larger surface for the photocatalysis. The degradation rate of OII reached 90.5 % during a period of 7 hours with Cu doped ZnO thin layers deposited on porous substrates modified with addition of $ZrO_2$.

## I. Introduction:

The contamination of water resources is a problem that is becoming acute today. This pollution results from the massive use of organic and mineral pollutants coming from agricultural, urban and industrial origin. Synthetic dyes used in the textile industry represent one of the most dangerous contaminants. Due to industrial manufacturings, the contamination is mainly provided by residue discharges into rivers. In order to preserve and improve the quality of waters surrounding factories at low costs for the benefit of the population, many treatment techniques for the purification of contaminated water are developed.

In this goal, the heterogeneous photocatalysis is a promising alternative for the treatment of organic pollutants in water. The discovery of the phenomenon of photocatalytic decomposition of water using samples of ceramics based on mullite and Zircon (a type of kaolinitic clay called DD3 made without and with a content of 38% of $ZrO_2$ treated at 1300 °C during 2h) and under UV irradiation has greatly contributed to this research field. The presence of zirconium oxide ($ZrO_2$) leads to the formation of zircon ($ZrSiO_4$) with a high rate of increasing porosity after consumption of silica in the vitreous phase. It has been stated [1] in a previous work that the addition of 38 wt% $ZrO_2$ to DD3 clay gives the highest rate of porosity (33%) on the ceramics surface. Most of the catalysts used in the current photodegradation are transition metal oxides. Among these oxides ones are the zinc oxide (ZnO) and the copper oxide (CuO). Their band gaps are 3.4 eV and 1.2 eV respectively [2-5]. They can be used in the photodegradation of organic compounds in aqueous media under UV irradiation [6], and even under visible light [7, 8]. However, these organic compounds have metastable phases and they must be stabilized before photocatalytic application.

The literature shows that oxides having high surface areas indicate a better photocatalytic capacity. Cu is then selected as an ideal doping for ZnO due to its similar physical and chemical properties [9-11], It can change the optical properties of the micro structure [13, 14] (band-gap decreasing, absorption shifted to longer wavelengths [12]) and change the morphology of ZnO nanostructures [10, 11].

In the aim of the synthesis of thin layers, we used in this work the autoclave method. Indeed, once the oxide layer is deposited, the role of ions for the absorption of impurities allows the development of photocatalyst and thereby leads to an optimizing efficiency.

Finally the final objective of this work is to study the effect of thin films of undoped ZnO (zinc oxide) and doped ZnO with different concentrations of copper deposited on ceramic substrates prepared by autoclave, on the photocatalytic activity of Orange II (OII) under UV light. ZnO has proved to be a good material for photocatalyst [15]. The obtained results show that a degradation of 90.5 % of OII by a thin layer deposited on porous ceramic substrates is reached after a time of 7 hours. This example of degradation of an organic compound can be easily extended to other damageable organic compounds. Moreover, this material offers a simple and low cost promising technique of fabrication for the degradation of organic compounds in a limited time and on large surfaces considering the large availability of the initial material.

## 2. Experimental study

### 2.1- Materials used

Two types of ceramics have been used in this work as primary materials based on a local DD3 clay. This material is kaolin obtained from the region of Guelma (Djebel Debagh), has a gray color and is provided by a ceramic company based in Guelma, Algeria [16]. The first one is a pure DD3 clay. Its three-mile structure exists naturally in the form of thin sheets and it is a group of quadrilateral layers with eight-faceted layers as shown in the figure 1 [17]. $SiO_2$ (> 40%) and $Al_2O_3$ (> 30%) with additional ions, are the most important compounds of the DD3 material [18]. The second material is a DD3 clay with 38% of $ZrO_2$.

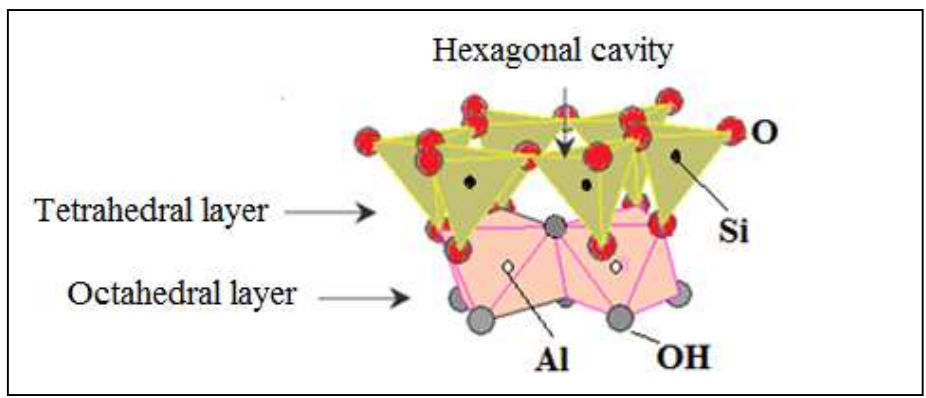

**Fig.1.a:** Crystal structure of kaolinite.

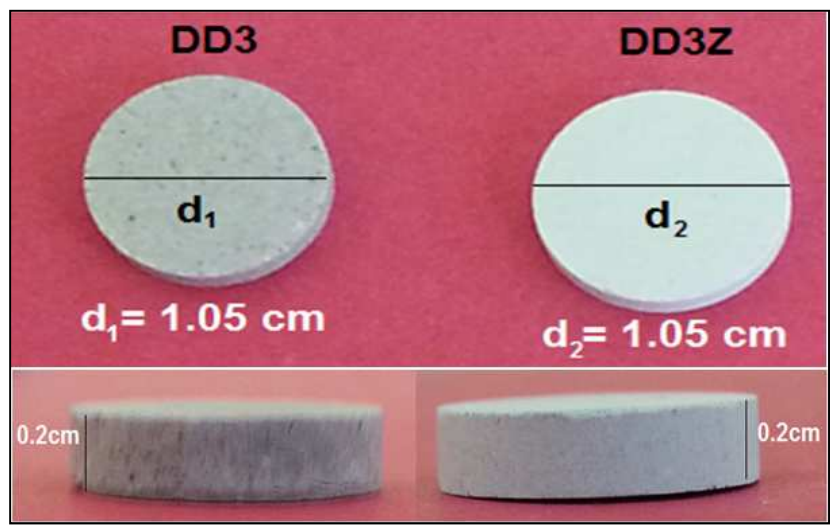

**Fig.1.b: Thickness and diameter of samples**.

For the mechanical properties of the two substrates, a Vickers hardness test was used for measuring the microscopic hardness using an 'AFFIRI' hardness tester. The test method consists in the indentation of the ceramic samples with a diamond indenter in the form of a right pyramid with a square base and an angle of 136 degrees between the opposite faces. We applied the following values for the load equal to 300, 500 and 1000 g. The area of the sloping surface after indentation is measured. The Vickers hardness is calculated by dividing the load value by the square area of indentation in millimeter. Each load was applied in four different regions of the two sample surfaces in order to calculate the mean area of the sloping surface after indentation.

| HV [Gpa] / Load [g] | DD3 [Gpa] | DD3+38% $ZrO_2$ [Gpa] |
|---|---|---|
| 300g | 4.49 | 1.53 |
| 500g | 6.11 | 2.71 |
| 1000g | 33.85 | 18.16 |

**Table 1.** The Vickers hardness (HV) measurements in SI units as a function of the load. The values vary according to the force applied.

The table 1 shows the change in the microscopic hardness in terms of the strength applied to the samples DD3 and DD3 + 38% $ZrO_2$ (DD3Z) treated at 1300°C. We note that the values of the microscopic hardness are increased by the applied force height, which is more pronounced when applying a load of 1 Kg. The low microscopic hardness for the DD3Z sample can be explained by the increase of the zirconium oxide which reduces the cohesion and convergence between the sample granules and hence increases the inside spaces. This is also explained by the quality of the phases generated by the heat treatment for both the cristobalite and the mullite of the DD3 type, which have good mechanical properties and begins to be reduced with the addition of zirconium oxide. We point out. that the formation of zircon ($ZrSiO_4$) induces a high rate of open porosity after consumption of silica in the vitreous phase.

**2.2. Characterization techniques**

In addition the following chemical products as zinc acetate [$((CH_3COO)_2 Zn.2H_2O)$], copper acetate [$(Cu(CH_3 COO)_2)$], absolute ethanol [$C_2H_5OH$], ethanolamine [$C_2H_7NO$] (MEA), and Orange II ($C_{16}H_{11}N_2NaO_4S$; OII) were used in this work for application.

The crystal structure of the particles was determined by X-ray diffraction (Bruker AXS-8D) with a Cu Kα(λ = 1.5406 Å) type of radiation. The surface morphology and the sample size were visualized by an atomic scanning electron microscope (JSM-6301F). An energy-dispersive X-ray spectrometer (EDX- X-Max 20 $mm^2$) was used for chemical composition. The spectra of the photocatalysis absorbance were measured by a UV-Vis spectrophotometer (V- 630, JASCO).

## 2.2- Preparation Methods
### 2.2.1- Preparation of the ZnO solution

In a 50 ml cleared beaker dried in an oven at 50ºC, three witness solutions of based Zinc acetate were prepared. These solutions were used for the autoclave method. A pure ZnO is formed when the zinc acetate (Zn $[OOCCH_3]_2 \cdot 2H_2O$) [19, 20] is dissolved in absolute ethanol. The solution gave us a molar concentration solution of 0.4 mol/l. Two contents of copper acetate (Cu $(CH_3 COO)_2$) with 4 wt% and 6 wt%, were added to two of the three solutions previously prepared. The obtained solutions were mixed using a magnetic stirrer during 2 minutes in a first step with a rotating speed of 500 rpm. 1:0 of catalyst ethanolamine ($C_2H_7NO$) was added. In a second step, we left the shaker during 2 hours at a temperature of 60 °C. The objective of this operation was to obtain a stable and homogeneous solution. During the agitation, the beaker was well covered to prevent any kind of contamination and evaporation of ethanol.

### 2.2.2-Preparation of substrates and thin films

After grinding and sifting the solutions, cylindrical samples (13 mm matrix) of clay DD3 and DD3 with 38% of $ZrO_2$ were made. The samples were treated at 1300 ° C during 2 h. The obtained ceramics were used as substrates for thin layers. The final deposited layers were active layers of ZnO and Cu doped ZnO. For this deposition we followed the thermal technique [21, 22]. The solution was prepared in the same manner as previously and deposited on the same type of ceramic (DD3, DD3Z) samples. First, we put 100 ml of distilled water in a first container (metal) and then put both the sample and the 100 ml solution in a second container (made from the flask) for the thermal preparation. After carefully closing the container, this one was placed over the magnetic mixer device at 300 °C during 75 minutes. We then deposited the layer above the ceramic sample. After this preparation, we let the sample dry at 200 ° C for 5 minutes. Finally a heat treatment at a temperature of 500 ° C during two hours was performed in order to obtain a crystalline surface. The various steps of this method are shown in the following diagram (Fig.2).

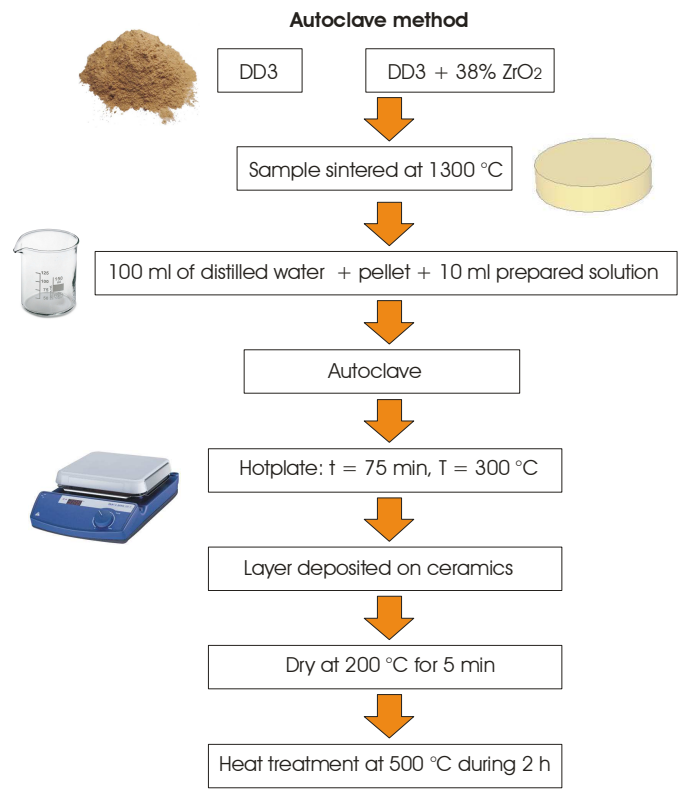

**Fig.2:** The method used in the preparation of thin layers.

## 2.3- Measurement of photocatalytic activity

The photocatalytic activity of all the fabricated materials was demonstrated by measuring their absorptions as a function of time. After the preparation of the aqueous solution of OII (12.5 mg/l; pH = 7.86), and the doped ZnO thin films deposited on ceramic substrates, 25 ml of an OII dye solution was put in direct contact with the samples . A fluorescent lamp (4W - 365 nm) was used for the activation of the reaction on the surface of the pellets. After each 1 h, 2.2 ml of the solution was taken in a quartz tube. The absorbance of the solution was analyzed by measuring the UV-Vis spectrum in the wavelength range of 250-650 nm. Distilled water was used as a reference.

After each photometric measurement, the reaction cell was put back to control the photocatalyst activity. The absorption rate of the degradation was calculated using the following equation [21, 22]:

$$\text{Dégradation ratio} = \frac{C_0 - C}{C_0} \times 100\%$$

where: $C_0$ and C are the initial concentration before and after illumination respectively.

# 3- Results and interpretations

## 3.1- X-ray diffraction (XRD)

The figure 3 presents the intrinsic characterization of the sample with the X-ray diffraction in the range $2\theta = 15 - 45°$. We show that the obtained results are the essential phases of ceramics: Mullite (JCPDS 15-0776), cristobalite (JCPDS 01-0424) for DD3-clay based substrates and Zircon (JCPDS 06-0266) and Zirconia (JCPDS 37-1484) for DD3+$ZrO_2$-clay based substrates. The characterization by X-ray diffraction when the thin film has been deposited shows the most intense peaks corresponding to zinc and copper oxide phases. New peaks appear with preferential orientations (100), (002) and (101) for the phase of the ZnO hexagonal wurtzite (JCPDS 03-0891). The effect of doping with 6 wt% Cu shows a single peak with a very high intensity (11-1) and (111). This orientation was observed for the diffraction of CuO and Cu phase (JCPDS 01-1117).

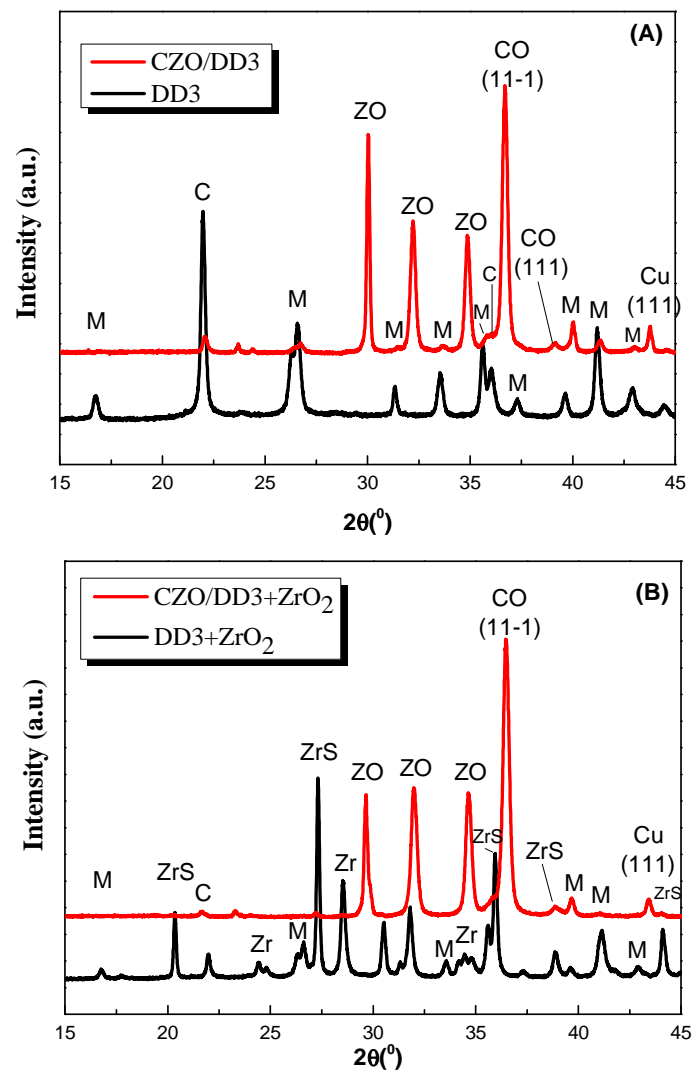

**Fig.3:** XRD spectra of ceramic pellets treated at 1300ºC prepared by the Autoclave method before and after doping. ZrS: Zircon (ZrSiO$_4$), Zr: Zirconia (ZrO$_2$), C: Cristobalite (SiO$_2$), M: mullite (3Al$_2$O$_3$.2SiO$_2$), ZO: zinc oxide (ZnO), CO: Copper oxide (CuO).

**3.2- Scanningelectron microscopy and energy-dispersive X-ray spectrometer**

The surface morphology analysis was studied with two selected samples corresponding to the two substrates: DD3 and DD3Z allowing a restriction to these samples for commodity of presentation.

The SEM image (Fig.4.a) of a DD3-clay substrate shows a near uniform morphology. However, the addition of ZrO$_2$ creates a more porous surface (Fig.4.b). After the deposition of thin films (Fig.5.e and f) the results show that the particles of ZnO and CuO precipitate, have almost a spherical shape and cover the surfaces of ceramics. Further experiments were done to characterize the elemental composition of samples by EDX spectrum (Fig.4.c, d and Fig.5.i, j). The figure (4.c, d) confirmed the presentation of all ceramics peaks such as Al, Si, Zr and O. One can see in the figure 5g and 5h, the characteristic signal of Zn, Cu and O after the deposition of the material (Cu doped ZnO (CZO)) with a disappearance of the basic elemental composition of ceramic substrates such as zirconium, silicium and aluminum. This result proved the formation of ZnO and CuO over ceramic substrates by calcination especially for the DD3+ZrO$_2$-clay. The average diameters of the grain size of these particles are 68 nm for the DD3-clay and 152 nm for the DD3+ZrO$_2$-clay (Fig. 5j and 5h).

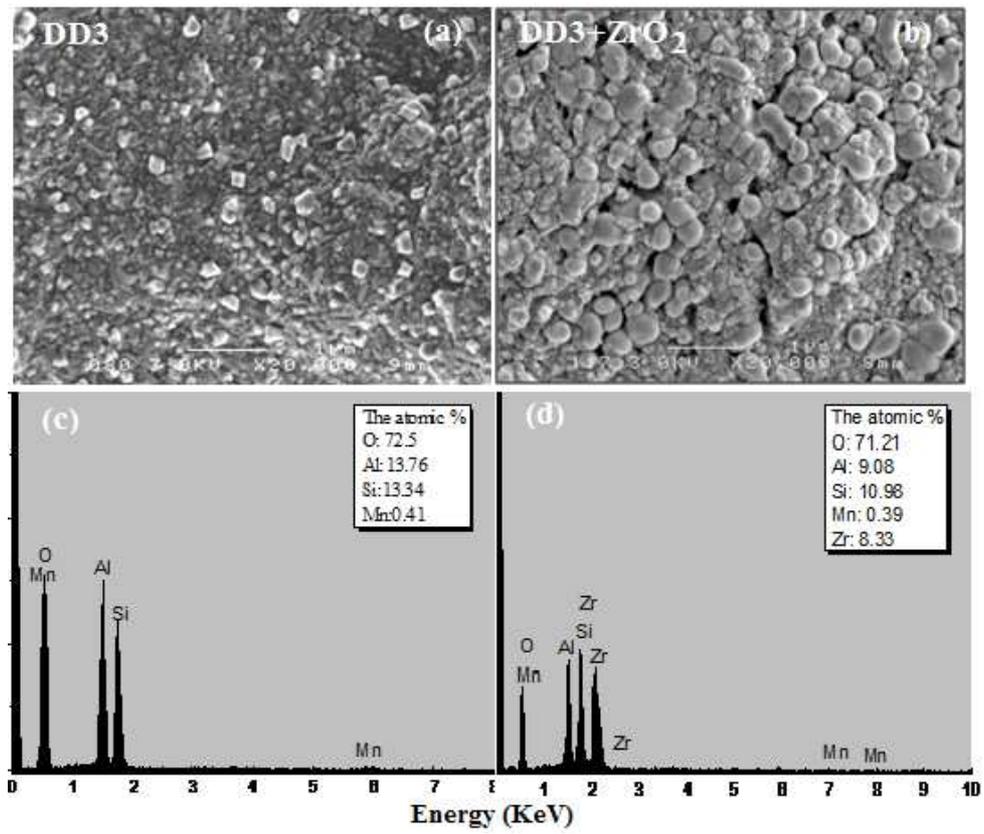

**Fig. 4:** SEM images and EDS spectrum of ceramic pellets before doping (a, c) DD3-clay and (b, d) after doping: DD3 + $ZrO_2$-clay. The scale bar is 1 µm.

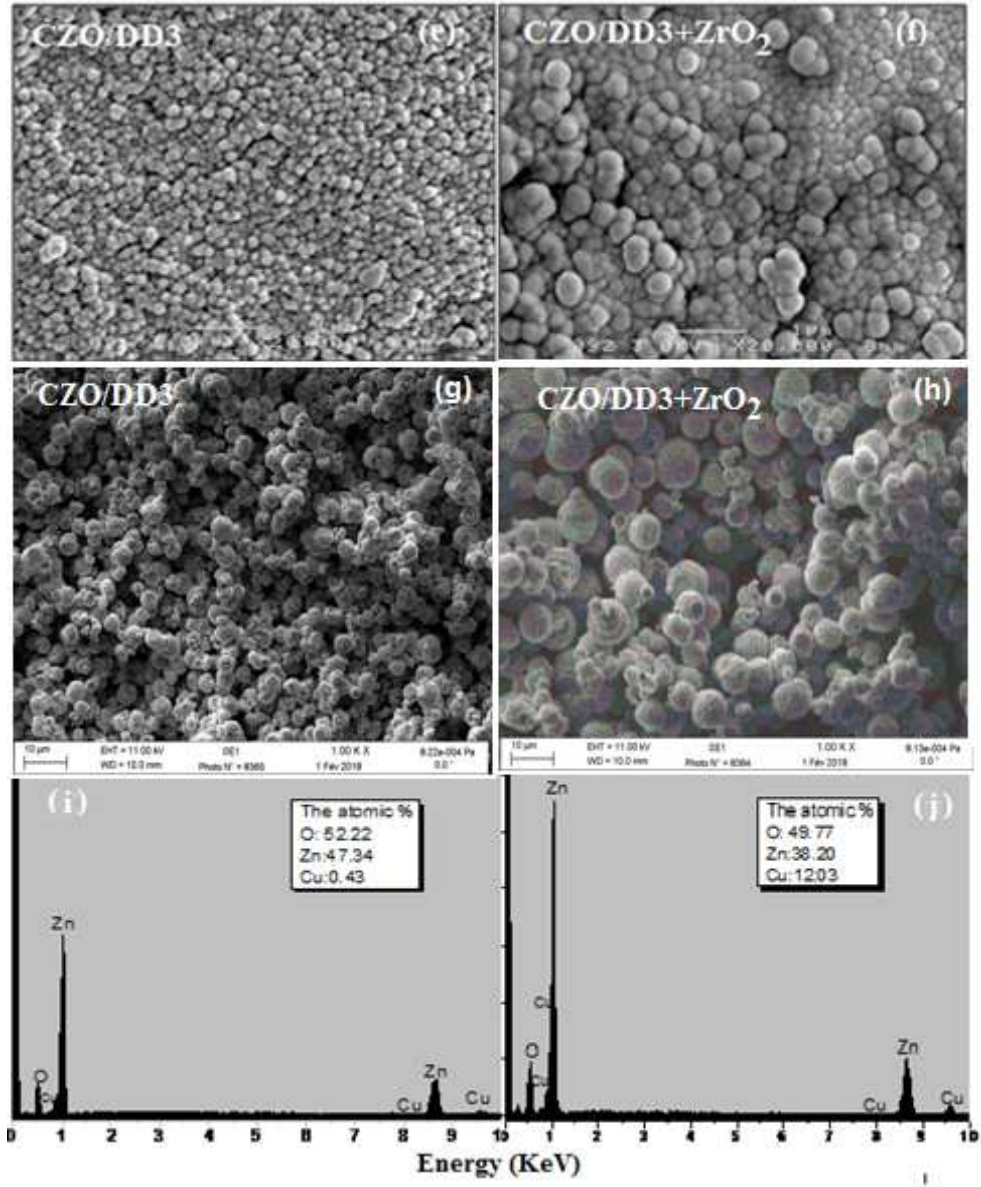

**Fig. 5:** SEM images and EDS spectrum of ceramic pellets after doping (e, i) DD3-clay (Cu doped ZnO: CZO) and (f, j) DD3 + $ZrO_2$-clay (Cu doped ZnO: CZO). (e, f) images with a scale bar of 1 μm and (g, h) images with a scale bar of 10 μm.

In order to determine the surface roughness of the images obtained with SEM (Scanning Electron Microscope) (Fig. 5), an appropriate measurement method was used: the height–height correlations (HHCF). This technique provides a quantitative and a single measurement of the roughness of the surface using the relationship between the HHCF function in terms of position r for the ceramic types before and after the deposition process as follows:

$$H(r) = 2\sigma^2 \left[ 1 - \exp(-(\frac{r}{\xi})^{2\alpha}) \right]$$

σ is the surface roughness, α is the surface roughness exponent and is used to calculate the fratal dimension (Hurst parameter), and ξ is the lateral correlation length which distinguishes the short-range and long-range behaviors of the rough surface. This is the lateral length at which the correlation function of the surface profile drops to $e^{-1}$ of its lateral length maximum.

From the figures 4 and 5 we calculate the value of all the dimensions mentioned above which are summarized in the table 1. We observe that even if the average surface roughness (σ) is almost identical in both substrates (DD3 and DD3Z), 122.5 nm and 123.1 nm respectively, the obtained Hurst parameter value (α) decreased from 0.81 to 0.75 when we added zirconium oxide. Values close to zero indicate a rough surface whereas values close to 1 indicate a smooth surface. α indicates a greater roughness on a very limited scale, which is strongly related to the holes on the substrate surface when we add $ZrO_2$.

After the Cu:ZnO layer deposition, a huge difference in the sample roughness is obtained. This is more than remarkable. The roughness σ reaches 123.5 nm for the substrates of DD3 and 130.8 nm for the substrate with $ZrO_2$. The correlation length of the samples (ξ), which represents the typical distance for a short-term behavior between two similar, long-term features of the coarse surface, is greater in the presence of zirconium oxide (44.3 nm for DD3 and 58 nm for DD3Z). This can be referred to the size of the granules that are larger for this type of sample according to the results of the SEM. When the active layers are deposited on ceramics, the Hurst coefficient decreases in both types of substrates and is estimated at 0.90 nm for the DD3 type and 0.76 nm when $ZrO_2$ is added, meaning that it has a high Hurst value and has a smoother surface. This indicates that spherical granular growth of the active layers is present in the substrates but is larger in size with the DD3Z substrate, which gives a higher roughness and therefore a wider surface area for an active photomodulation. This indicates that microscopic holes have an important role to play in the process of crystal growth when sedimentation is applied by immersing the hydrothermal method. The larger Hurst coefficient was for the surface area of the smooth samples. It was confirmed by calculating the average size of the granules. For DD3 pellet the particle size is 154.94 with an active layer and it was 198.587 nm without the active layer. It reached 338.75 nm 331.148 nm for DD3Z substrate with and without the active layer respectively. So the active layer modificates and increases the grain size.

| *Substrates* | *DD3-Clays* | | *DD3+38%ZrO$_2$-clays* | |
|---|---|---|---|---|
| *Samples* | *DD3* | *CZO-DD3* | *DD3Z* | *CZO-DD3Z* |

| σ(*nm*) | 122.5 | 123.5 | 123.1 | 130.8 |
|---|---|---|---|---|
| ξ(*nm*) | 44.3 | 75.5 | 58 | 57.9 |
| A | 0.81 | 0.90 | 0.75 | 0.756 |
| Grain size (nm) | 198.587 | 154.94 | 331.148 | 338.75 |

**Table.1.** Calculated σ, ξ, α and grain size of ZnO:Cu layer deposited on the different substrates

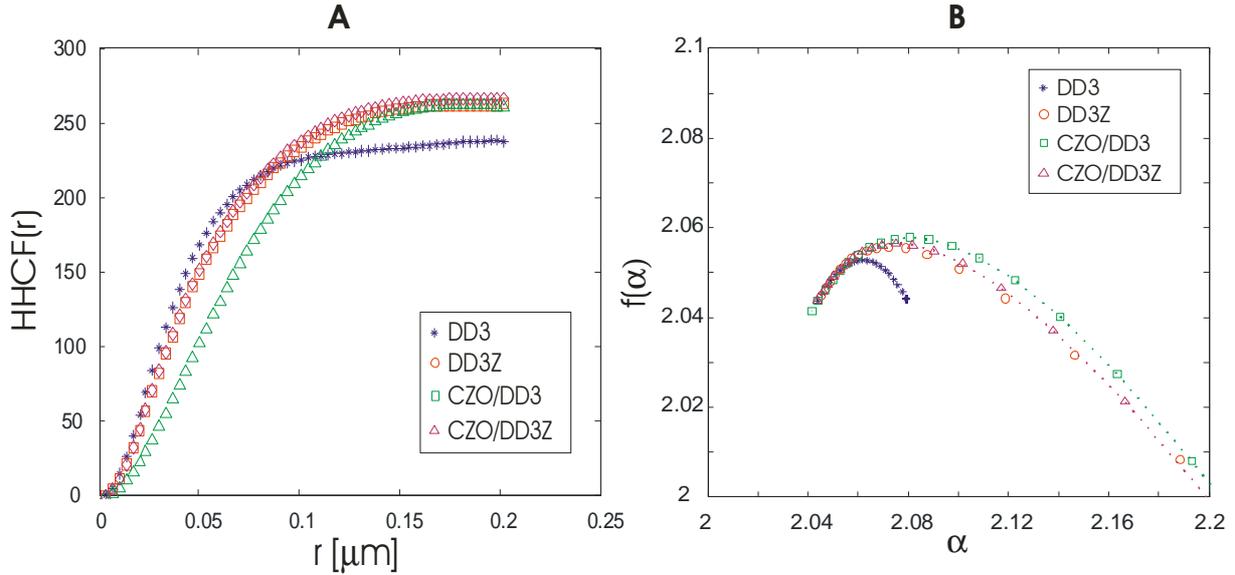

**Fig. 6.** a) Height–height correlation functions for comparison of substrates of DD3 and DD3 with ZrO2 (DD3Z) with ZnO:Cu layers deposited on DD3 and DD3+ZrO2 substrates (CZO/DD3 and CZO/DD3Z respectively), b) Multifractal spectrum of the different samples.

We calculate the multifractal spectrum f(α) which gives an information on the surface complexity (Fig.6) [24]. The multifractal spectrum shows the distribution of scaling exponents for a surface and additionally the spectrum width (fig. 6b) provides a measure of how much the local regularity of a roughness varies in the two spatial directions. The wider is the singularity spectrum, the higher is the heterogeneity. Local scaling indices of the surface roughness measure an increased non-uniformity of the height probabilities of the surface and vice versa. A surface that is monofractal exhibits the same regularity everywhere on the surface and therefore has a multifractal spectrum with a narrow spectrum. The symmetry/asymmetry of the singularity spectrum provides information about the structure homogeneity/heterogeneity.

We see clearly the difference between the multifractal spectrum of DD3 and the multifractal spectra of the other modified DD3 based materials where the spectrum is broadened with non uniform structures on the surface.

### 3.3UV-Vis

### 3.3. UV–visible spectra – optical properties

UV–visible optical absorption spectra of the undoped ZnO and doped ZnO have been carried out for different Cu concentrations from 0% to 6 % deposed on the Ceramic substrate at room temperature using a UV– visible spectrometer in the range of 200 nm -900 nm (Fig.7). For DD3Z-clay, the small absorption band is observed in the range of 200–290 nm corresponding to tetragonal or cubic structure of $ZrO_2$ [25, 26]. The absorbance onset is at about 385nm for bare ZnO particles. Furthermore, a broad absorption is seen around 681–683 nm and might be attributed to the presence of $Cu^{2+}$ species in the prepared Cu:ZnO thin films [27].

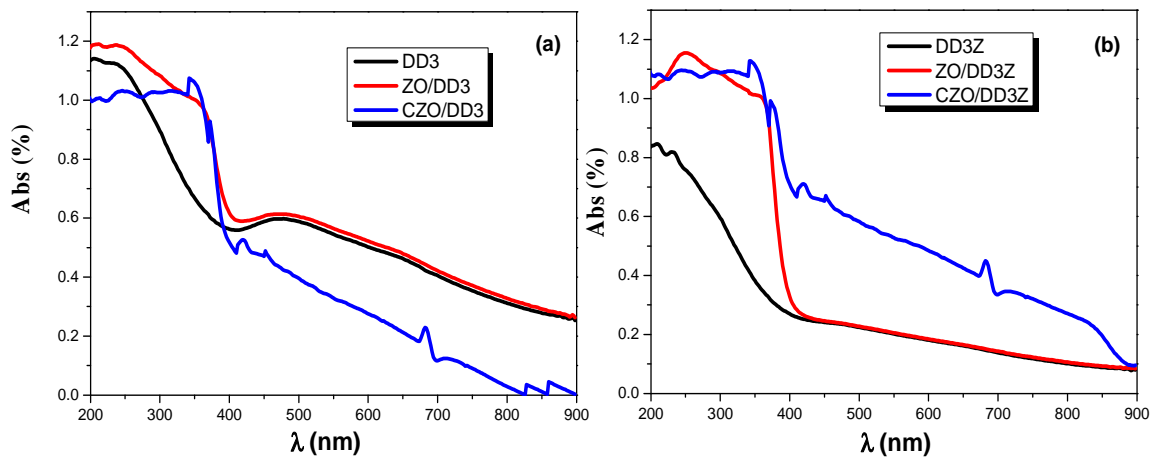

**Fig. 7.** UV-Visible Spectra of undoped and Cu-doped ZnO thin layer deposed on pellets substrate (a) DD3-clay and (b) DD3+38 wt% $ZrO_2$-clay

The main absorption maximum of undoped and doped nanostructures is high and rather sharp on DD3-clay. When we use DD3+38 wt% $ZrO_2$-clay the absorption is reduced. Comparing to the undoped nanostructures, doping with Cu increases the main absorption on DD3Z clay.

### IR spectra:

The chemical compositions of the samples were examined by FTIR spectroscopy in the region 400 - 3000 cm$^{-1}$ and the observed result is demonstrated in the figure 8a and 8b. The two types of the ceramic pellets, before and after the deposition of the ZnO layer and Cu:ZnO

(DD3 and DD3Z), show the same IR characteristic. The difference between the two characteristics is only in the confined area 400-1250 cm$^{-1}$. We also note that when the doping increased, the intensity of the peaks are increasing and appear strongly, especially for the zinc deposition above the sample which has a high percentage of porosity (DD3+ZrO$_2$). These results showing the presence of greater doped proportions compared to the other undoped type (DD3) are confirmed by the EDX analyses. The IR spectrum exhibited several well-defined absorption bands appearing at 420, 470,480, 490, 508,664, 1009, 1649 and 2368 cm$^{-1}$. Two bands characteristic of the mullite (Si-O-Al) and cristobalite (SiO$_2$) formation are observed at 840 and 526 cm$^{-1}$, respectively [28]. The addition of ZrO$_2$ on the DD3-cly shows two peaks at 1009 cm$^{-1}$ and one peak at 490 cm$^{-1}$ indicating the formation of Zircon (ZrSiO$_4$) characterized by m-ZrO$_2$. The band that appears at 664 cm$^{-1}$ 2368 cm$^{-1}$ is assigned to the vibration of the Al-O octahedral bond indicating the development of Al$_2$O$_3$ in the structure. The last band at about 1649 cm$^{-1}$ is assigned to the bending and stretching vibrations of the O–H bond due to adsorbed water molecules. The sample showed a band with the peak centered at 420 and 480 cm$^{-1}$ correlated to zinc oxide bond which corresponds to the Zn-O vibration. For the doping with cooper there is two sharp intense peaks observed at 470 and 508 cm$^{-1}$ for the Cu-O vibration.

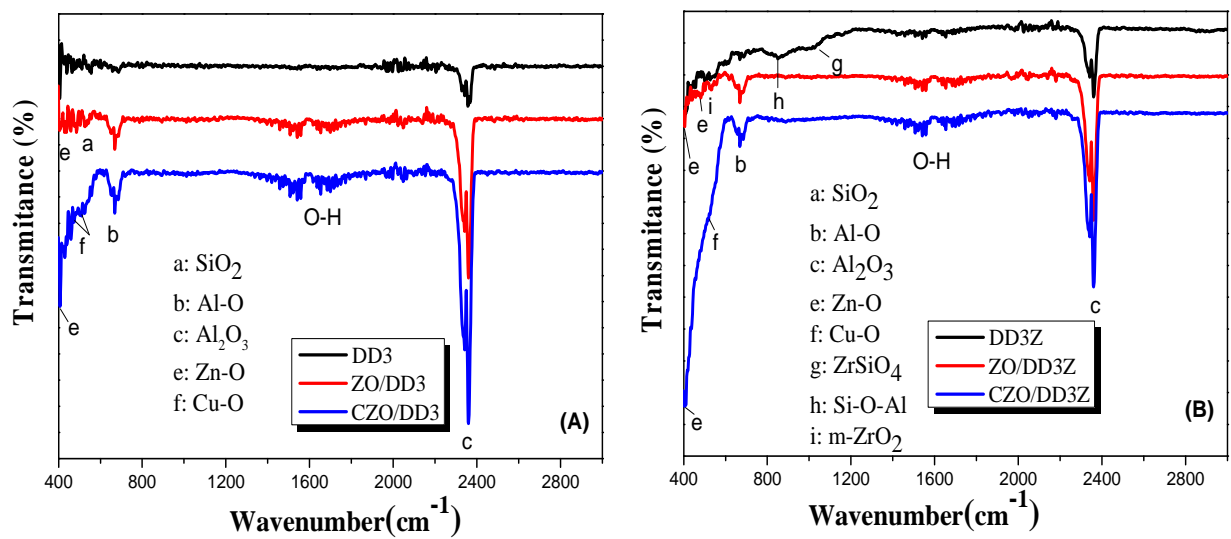

**Fig. 8.** Infrared Spectra of pellets (A) DD3-clay and (b) DD3+38 wt% ZrO$_2$

### 3.4- Confocal spectra

The study of the influence of Cu doping ZnO on the luminescence of DD3 and DD3Z-clay substrates was done with measured confocal spectra presented in the figure 9. When the sample is doped with 6 % of Cu atoms, our data analysis indicates that the confocal emission intensity decreases quickly with a further increase of the Cu dopant. Furthermore, the decrease of the emission intensity with the increase of the cu concentration can be assigned to the increase of the non-radiative relaxation which is due to the surface bound states introduced by copper ions and the degradation of ZnO crystal quality [29-31]. Also this increase may be attributed to the increase of the lattice defects, resulting in a good agreement with the recently reported works in the literature [32]. This study was experimented for two main reasons: to study i) the quantum size effect and ii) the structural defects of the deposited layer. When the ZnO and Cu:ZnO layer is deposited, the emission band is broadened and it can be speculated that the band width of the ceramic is widened owing to the replacement by the $Zn^{2+}$ and $Cu^{2+}$ [33 - 34]. The near band edge (NBE) emission at about 649 nm (DD3) is attributed to ZnO and can be observed in the spectra (Fig. 9a and b).

This additive compound prevents a recombination (e-/h+) and creates many oxygen vacancies which increase the copper doping. Generally, the photocatalytic activity is related to the particle size and concentration of oxygen vacancies on the surface. Hence, it is believed that the prepared Cu:ZnO nano-crystals with a big particle size, a high surface area and high surface defects would enhance the photocatalytic activity of pellets with active layers.

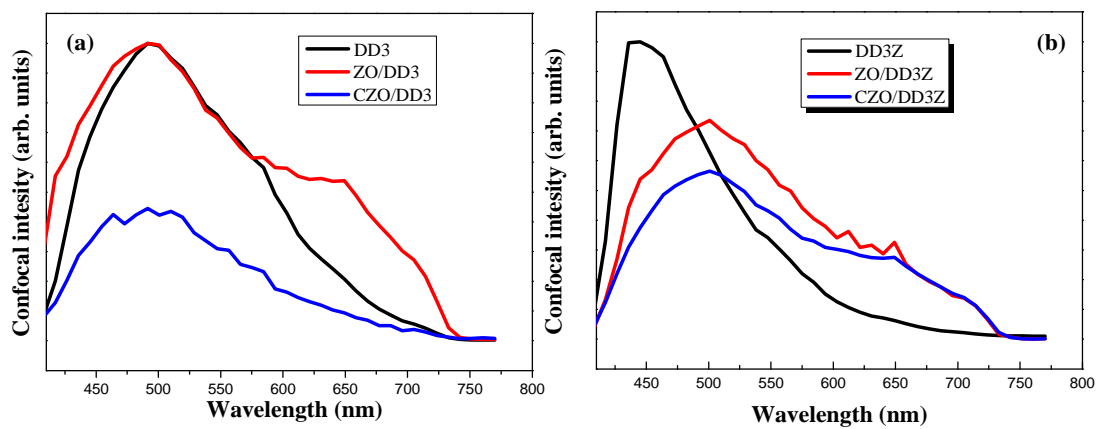

**Fig 9.** Confocal Spectra of undoped (DD3), ZnO doped (ZO DD3) and Cu-doped ZnO (CZO DD3) thin layer deposed on pellets substrate (a) DD3-clay and (b) DD3+38 wt% $ZrO_2$-clay

### 3.5- photocatalytic performance

In order to demonstrate the possibilities for practical applications of the prepared samples, we studied the use of our samples for the purification of a contaminated solution. This study is carried out by measuring the degradation of Orange II as a function of time. The UV-visible spectra obtained for thin layers prepared by the autoclave method are shown in the Figure 11. These spectra are typical of those of aqueous solutions of OII, with a maximum absorbance at 484 nm. The layers are deposited on ceramic based DD3-clay and DD3 + $ZrO_2$-clay substrates. In these two figures it is noted that the degradation for short periods is almost negligible for all the samples. Considering the samples of DD3 + $ZrO_2$ with ZnO thin layers doped by 6% of Cu, a suitable effect on the purification of Orange II reaches 81.1 6% for a degradation time of 7 hours. However, the photocatalytic activities of thin layers deposed on DD3 substrate with the same doping and the period of dye degradation did not exceed 36.10% (fig.12). Cu:ZnO on DD3Z has the best performances for the degradation of OII. It is 4.5 more efficient than the other sample. This important effect can be connected to the important role of the open porosity of the samples based on DD3+ $ZrO_2$-clay compared with the DD3-clay, when the filling with particles of new phases (ZnO, CuO) precipitated after the doping. This is shown in Table 2 where the pore sizes decreases after sedimentation.

**Table.2.** Calculated pore size of substrates before and after deposited Cu:ZnO layer.

| *Substrates* | *DD3-Clays* | | *DD3+38%ZrO$_2$-clays* | |
|---|---|---|---|---|
| *Samples* | *DD3* | *CZO-DD3* | *DD3Z* | *CZO-DD3Z* |
| **Pores size (nm)** | 102 | 74 | 195 | 161 |

Semi-permeable ceramic membranes covered by an active layer (Cu:ZnO), whose pore diameters is between 102 and 195 nm, have been used for the separation of elements contained in a liquid. This technique is therefore used for the treatment of polluted water and effluents. The ultrafiltration that is done is mainly used to separate dissolved particles and degraded the impurities of the OII. In order to see the spread of colored liquid (OII) inside the samples and to know the time required, we placed a drop on each sample (before and after the deposition process) and took pictures of the stages of this spread of the upper and lower face of the pallets with the diameter of the effect of the droplet. A Vernier caliper was used to measure both diameters and thicknesses. The table (3) shows the results.

**Table.3.** Mass, diameter and thickness of all the pellets.

| *Samples* | DD3 | DD3Z | CZO/DD3 | CZO/DD3Z |
|---|---|---|---|---|
| mass (g) | 0.4053 | 0.4744 | 0.4583 | 0.5888 |
| diameter (cm) | 1.05 | 1.05 | 1.09 | 1.065 |
| thickness (cm) | 0.2 | 0.2 | 0.21 | 0.235 |
| d1 (droplet effect of the upper face) (cm) | 0.64 | 0.68 | 0.66 | 0.73 |

| d2 (droplet effect of the lower face) | 0.41 | / | / | / |

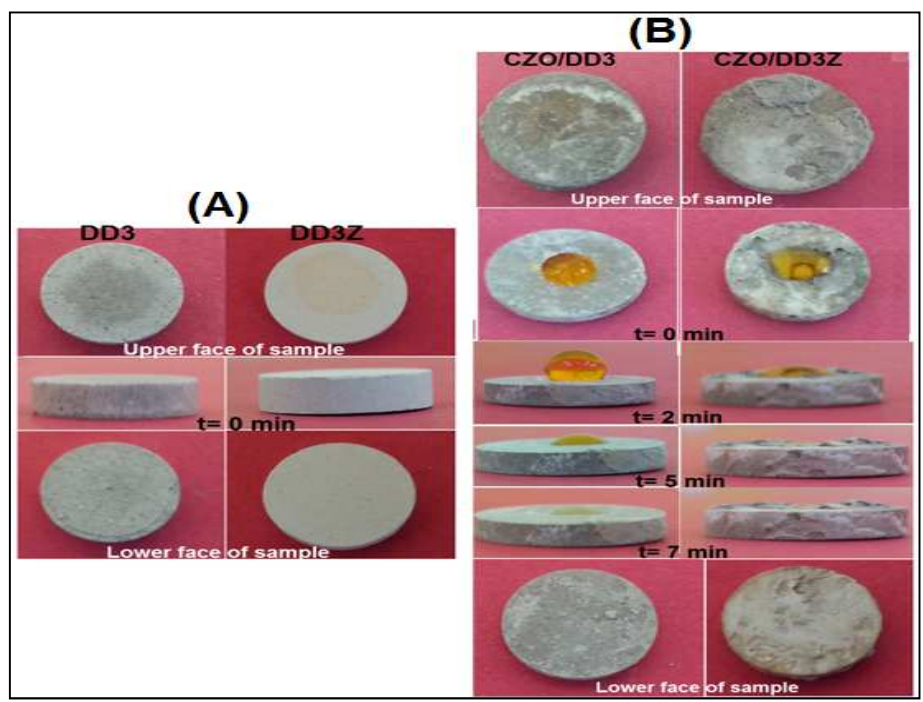

**Fig.10.** Images illustrate the spread of OII drop within samples.

After layer deposition with 6 wt% Cu doped zinc oxide (CZO), the open pores are filled with an additional amount which covers all the surface of the pellets and thus leads to an increased deposit (ZnO, CuO). The existence of these two oxides leads to the degradation of the OII organic dye in a more efficient manner. It is noticeable that there are pores or spaces between the granules precipitated after the deposition process, which allows the solution to penetrate inside the layer and to reach the surface of the sample (fig.13). For more understanding in order to see the role that the pores play during the degradation process the figure 13 shows the ratio of the degradation of OII as a function of the pore sizes after 5 minutes of the process,.

Moreover, the pH is also an important operational variable during the water treatment. In photocatalytic degradation systems, the pH value is one of the factors that influences the rate of degradation. The table 4 shows the changes for this value after this photocatalytic operation. Generally, when a compound is partially ionized or carrying charged functions, it is necessary to consider the electrostatic interactions that may take place between pellets and this compound.

The pH is equal to 8.64. It appears from these results that photo-decoloration is more important when the pH is basic pH. The effect of the initial pH of the solution on the photodegradation process is very important because it influences the photocatalyst characteristics, such as the electric charge of the particles, the size of the formed aggregates, the position of the conduction and valence bands of the semiconductors, catalysts, as well as the degree of adsorption of electron donors on their surfaces.

**Table.4.** Effect of pH values on the degradation of Orange II for the various pellets.

| PH of OII  | DD3  | DD3Z | CZO/DD3 | CZO/DD3Z |
|------------|------|------|---------|----------|
| initial PH | 7.86 |      |         |          |
| finale PH  | 7.56 | 7.80 | 7.98    | 8.64     |

The process of photocatalytic degradation of orange II with Cu:ZnO deposited on ceramics substrate for photocatalysis can be described as follows: i) the first step involves adsorption of the dye onto the surface of sample. ii) exposure of dye adsorbed on the Cu:ZnO surface by the UV light leading to the generation of electron-hole ($e^-/h^+$) pairs in ZnO and CuO [35-37]. The photogenerated electrons in the conduction band of ZnO and CuO interact with the oxygen molecules adsorbed on surface substrate to form superoxide anion radicals ($•O^{2-}$). iii) The pores generated by the valence band of ZnO and CuO react with surface hydroxyl groups to produce highly reactive hydroxyl radicals (•OH). These photogenerated holes can lead to dissociation of water molecules in the aqueous solution, producing radicals. The highly reactive hydroxyl radicals (•OH) and superoxide radicals ($•O^{2-}$) can react with the OII dye adsorbed on surface structures and lead to its degradation. The photocatalyse process is then based on the creation of electron-holes ($e^-/h^+$) leading to the generation of radicals (• OH, $•O_2^-$) which reduce OII. Cu allows to enhance the activity of ZnO by creating a larger number of electrons and so a bigger possibility to create more radicals $•O_2^-$ giving a white color to the contaminated solution. Moreover, the generated holes interact with $H_2O$ and $OH^-$ in order to create more • OH (the main oxydant) leading to the degradation of the dye in solution.

The studied samples (Fig.5e and f) proved that the specific surfaces with particles growing on the DD3+$ZrO_2$-clay substrate are more effective for degradation of OII than on a DD3 substrate.

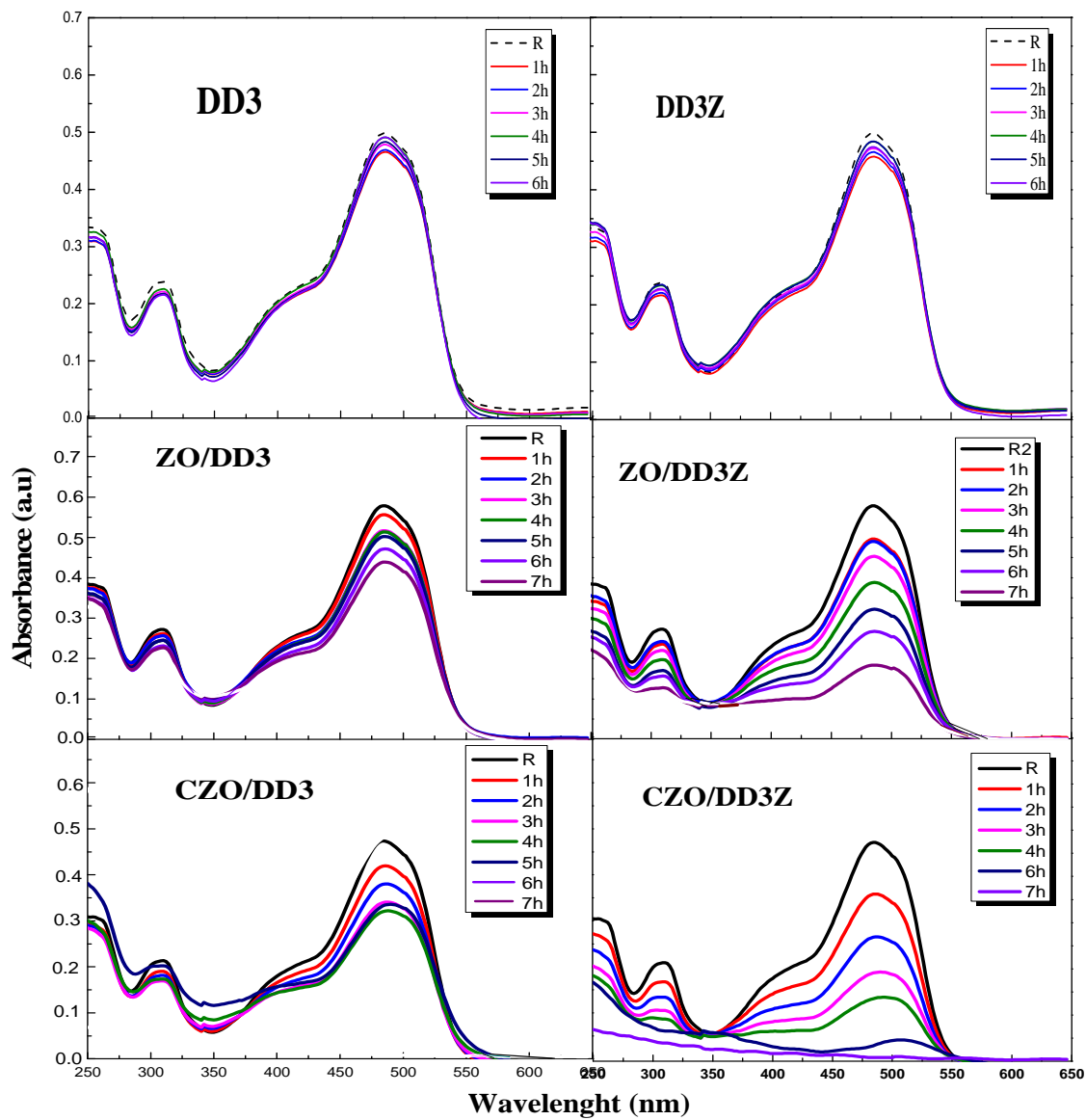

**Fig.11.** The degradation of OII by pellets of DD3-clay and DD3 + ZrO$_2$-clay fabricated with the thermal method.

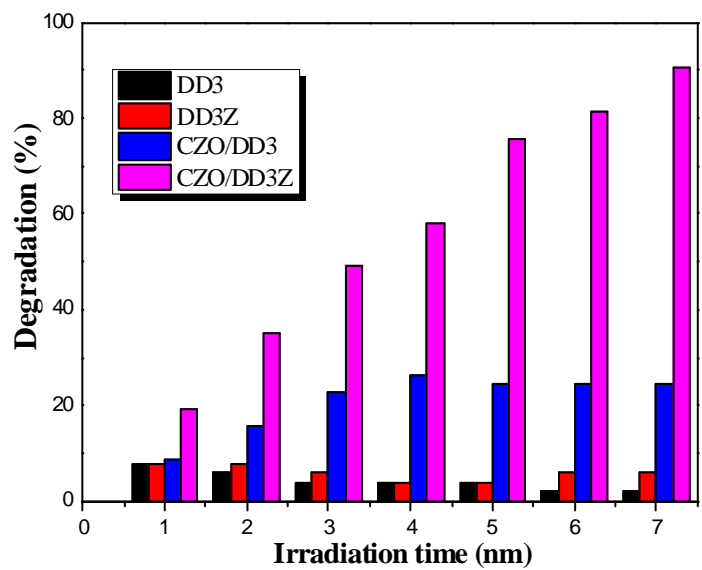

**Fig.12:** Degradation of orange II with different Cu-doping concentrations

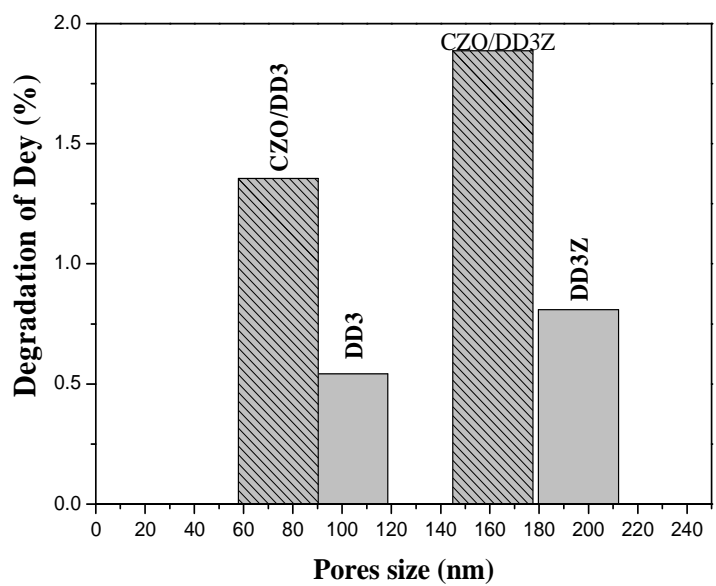

**Fig.13:** The ratio of degradation of OII in terms of pores size after 5 min.

## Conclusion

In this study, we have studied two different based kaolinite substrates. This is the most abundant phyllosilicate mineral which can be used as absorbent for removing water pollutants. We use in our study kaolin from Djebel Debagh 'DD3' doped with $ZrO_2$ and doped with several Cu-doped ZnO thin layers prepared by an autoclave method. The effects of this doping on the structural activity of photocathalysis are studied. The morphology of the layers is characterized by MEB; it was found that an addition of 6% Cu doped ZnO layer on the ceramic substrates (DD3 and DD3 + ZrO$_2$) increases the surface energy resulting in the formation of spherical-like particles. The EDX, IR and confocal analyses confirmed the presence of Zn, Cu and O as essential elements dominating after deposition when they cover all the surface of ceramic substrates. The photocatalytic activity of pellets is performed with the degradation of Orange II dye by measuring UV-Visible spectra as a function of the pore sizes. The results present a great effect on the purification with thin layers of porous ceramic substrates (DD3+ZrO$_2$-clay).

**Acknowledgment**

This work has been supported by the Laboratory of Active Components and Materials (LACM) of Larbi Ben M'hidi University - Oum El Bouaghi, Algeria and the laboratory of MOLTECH-Anjou, University of Angers, France.